\documentclass[twocolumn,a4paper]{article}
\usepackage{epsfig}

\input tcilatex
\begin{document}

\twocolumn[

\begin{center}

\textbf{\Large Intrinsic tunneling spectroscopy in small Bi2212 mesas}

\vskip1ex { Vladimir Krasnov$^{1,2}$, August Yurgens$^{1,3}$, Dag Winkler$^{4,1}$ and Per
Delsing$^1$ }

\vskip0.5ex \textsl{ $^1$ MINA, Chalmers University of Technology,
S41296, G\"oteborg, Sweden\\ $^2$ Institute of Solid state
Physics, 142432 Chernogolovka, Russia\\ $^3$ P.L.Kapitsa Institute, 117334 Moscow, Russia\\
$^4$ IMEGO Institute, Aschebergsgatan 46, S41133, G\"oteborg, Sweden}

\end{center}
]

%\textbf{Abstract:} We present experimental study of small Bi$_2$Sr$_2$CaCu$_2 $O$_{8+x}$
%mesa structures, containing few intrinsic Josephson junctions.
%The mesas exhibit clear tunnel-type current-voltage characteristics. This
%allows us to distinguish and simultaneously observe superconducting and
%pseudo gaps. We show that the superconducting gap vanishes at the critical
%temperature, while the pseudo-gap may sustain up to room temperature.\\

Tunneling spectroscopy of high-$T_c$ superconductors (HTSC)
provides an important information about quasiparticle density of
states (DOS), which is crucial for understanding HTSC mechanism.
Surface tunneling experiments \cite{Renner,Wilde} showed, that
besides the superconducting gap in the DOS, $ \Delta $, there is a structure
usually referred to as the ''pseudo-gap'', which exists well above $T_c $. Both the
behaviour of the superconducting gap and it's correlation with the
pseudo-gap are still a matter of controversy\cite{Renner,Wilde}.

To avoid drawbacks of surface tunneling experiments, such as
dependence on the surface deterioration, surface states and undefined
geometry\cite{Wilde}, we used ''intrinsic'' tunneling
spectroscopy. HTSC single crystals can be considered as stacks of
atomic scale intrinsic Josephson junctions (IJJ's). Using
microfabrication, it is possible to make small HTSC mesa
structures with a well defined geometry\cite{Yurgens}. Moreover,
IJJ's far from the sample surface can be measured, and
deterioration of the sample surface becomes less important.
Current-voltage characteristics (IVC's) of mesas exhibiting tunnel
junction behavior and can be used for studying
DOS\cite{Suzuki}. On the other hand, intrinsic tunneling spectroscopy can suffer from
ohmic heating of the IJJ's and steps (defects)
on the surface of the crystal.

In this paper we present experimental data for small area Bi$_2$Sr$_2$CaCu$%
_2 $O$_{8+x}$ (Bi2212) mesas containing few IJJ's. By decreasing
the mesa area, $S$, we minimize both the effect of overheating and
the probability of defects in the mesa. Mesas with dimensions from
2 to 20 $\mu $m were fabricated simultaneously on top of Bi2212
single crystals. First a long and narrow mesa was fabricated using
photolithography and chemical etching. Next, insulating CaF$_2$
layer was deposited and lift-off was used to make an opening.
Finally, Ag film was deposited and electrodes were formed on top
of the initial mesa by photolithography and Ar-ion etching. After
etching, mesas beneath Ag electrodes remain.

In Fig.1, normalized IVC's of three mesas at $T$=4.2K and $T$=150
K are shown. The vertical axis represents the current density,
$J=I/S$, and the horizontal axis shows the voltage per junction,
$V/N$. IVC's for different samples are plotted by
different colours. Parameters of the mesas are listed in Table 1.
From Fig.1 it is seen, that the normalized IVC's merge quite well
into one curve. This indicates good reproducibility of the
fabrication procedure. The $c$-axis normal resistivity was, $\rho
_N=R_NS/Ns=44\pm 2.0$ $\Omega $cm, where $s=15.5\AA $ is the
spacing periodicity of IJJ's and $R_N$ is the resistance at large bias current and $%
T_c\simeq 93$ K.

\begin{tabular}{lllll}
Table 1: & mesa & $S$ ($\mu $m$^2$) & $N$ & $R_N (\Omega)$ \\
& S251b & $6\times 6$ & 12 & 229.9 \\
& S255b & $5.5\times 6$ & 12 & 256.4 \\
& S211b & $4\times 7.5$ & 12 & 272.5 \\
& S216b & $4\times 20$ & 10 & 87
\end{tabular}

From Fig.1 it is seen that IVC's exhibit clear tunnel junction behavior: (i)
At low bias, multiple quasiparticle branches are seen, representing
one-by-one switching of IJJ's into the resistive state\cite{Yurgens,Suzuki}.
The number of IJJ's in the mesa was obtained by counting those branches.
(ii) At intermediate currents there is a pronounced knee in IVC's,
representing the sum-gap voltage, $V_g=2\Delta /e$. (iii) At high currents,
there is a well defined normal resistance branch, $R_N$. As seen from Fig.1,
$R_N$ is almost temperature independent, as may be expected for a tunnel
resistance. In contrast, the zero bias resistance, $R(0)$, increases sharply
with decreasing $T$, as shown in inset b).

Inset a) in Fig.1 shows temperature dependence of the gap voltage, $V_g$,
and the maximum spacing between multiple quasiparticle branches, $\Delta V$,
for four mesas on two diffrerent chips. Solid and open symbols represent branches for $V>0$ and $V<0$%
, respectively. $\Delta V$ was determined at the ''maximum
critical current'', at which the last IJJ switches to the
resistive state, e.g. in Fig.1 that would correspond to $J\simeq
1.3\times 10^3$ A/cm$^2$. This current is fluctuating from run to
run, therefore, causing an uncertainty in determination of $\Delta
V$. From the inset a) it is seen, that $\Delta V$ is approximately
two times less than $V_g$. This is simply due to the fact that all
the IJJ's switch to the resistive state before they reach the gap
voltage, see Fig.1. However, it is seen that the temperature dependence of $%
\Delta V$ reflects that for $V_g$. The observed $V_g$ correspond
to $\Delta \simeq 32$ meV at $T$=4.2 K, in agreement with
\cite{Wilde}.

\begin{figure}[hbt]
\centering \epsfig{file=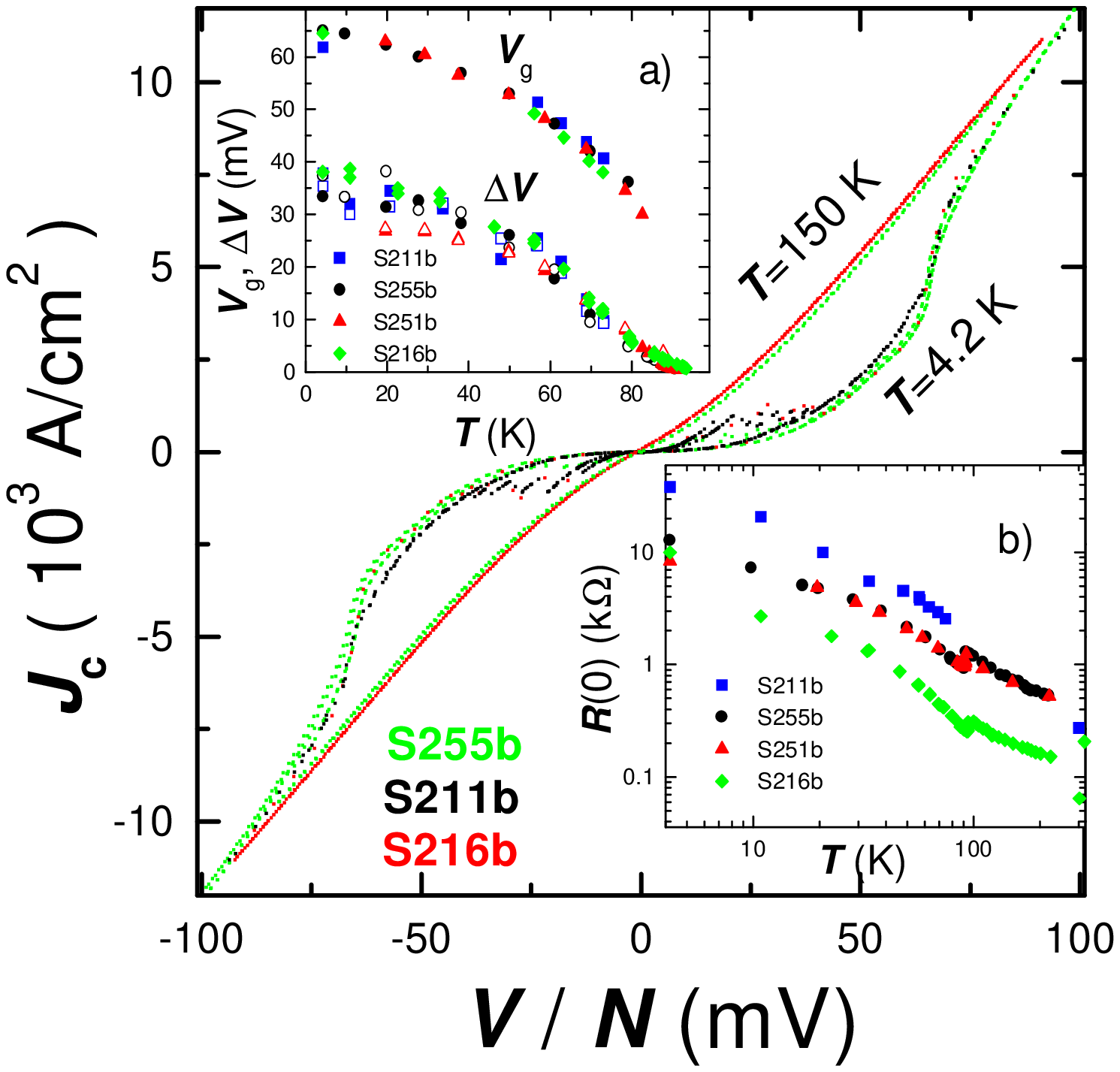,width=0.9\columnwidth}
\caption{$J$ vs. $V/N$ curves for three mesas at $T$=4.2K and
$T$=150 K. Inset a) shows temperature dependencies of the gap
voltage, $V_g$, and the voltage separation of multiple
quasiparticle branches, $\Delta V$. Inset b) shows temperature
dependencies of the zero-bias resistance.} \label{fig1}
\end{figure}

The IVC's in Fig.1 suggest that there is no significant heating
effect in our mesas. Indeed, overheating should cause the
reduction of $V_g$. Instead, we observed that the normalized IVC's
merge into a single curve with
identical $V_g$, despite a considerable difference in the dissipated power, $%
P\propto S$. Moreover, we have checked that the voltages of the
quasiparticle branches scale with their number, $V_n\simeq \frac
nNV_N$, where $V_N$ is the top branch, with all IJJ's in the
resistive state. Thus, switching of additional IJJ's does not
cause visible overheating of the mesa. On the other hand, we did
observe a strong overheating for even
smaller mesas (2$\times 4$ $\mu $m$^2$), containing larger number of IJJ's $%
(\sim 100)$, so that a clear back-banding was seen at large
currents, and significantly lower $V_g$ was obtained, similar to
that in Refs.\cite {Yurgens,Latysh}. Therefore, a small number of
IJJ's in the mesa decreases the risk of overheating.

In Fig.2, the conductance at different temperatures is shown for
one of the samples. The sharp peak at $V_g$ and the depletion of
conductance at $\left| V\right| <V_g$ is seen at low $T$,
representing the superconducting gap in the DOS. The suppression
of DOS below the gap results in strong temperature dependence of
$R(0)$, see inset b) in Fig.1. With increasing temperature, the
peak at $V_g$ shifts to lower voltages and decreases in magnitude.
At $T\sim $80 K, the peak is smeared out completely and only
smooth
depletion of the conductance remains at $V$=0. With the further increase of $%
T$, this depletion gradually decreases, but is still visible even at room
temperature, representing the pseudo-gap in DOS. In agreement with the
surface tunneling experiments\cite{Renner,Wilde}, there is almost no changes
in the conductance at $T_c$, which implies that the pseudo-gap coexists with
the superconductivity.

\begin{figure}[hbt]
\centering \epsfig{file=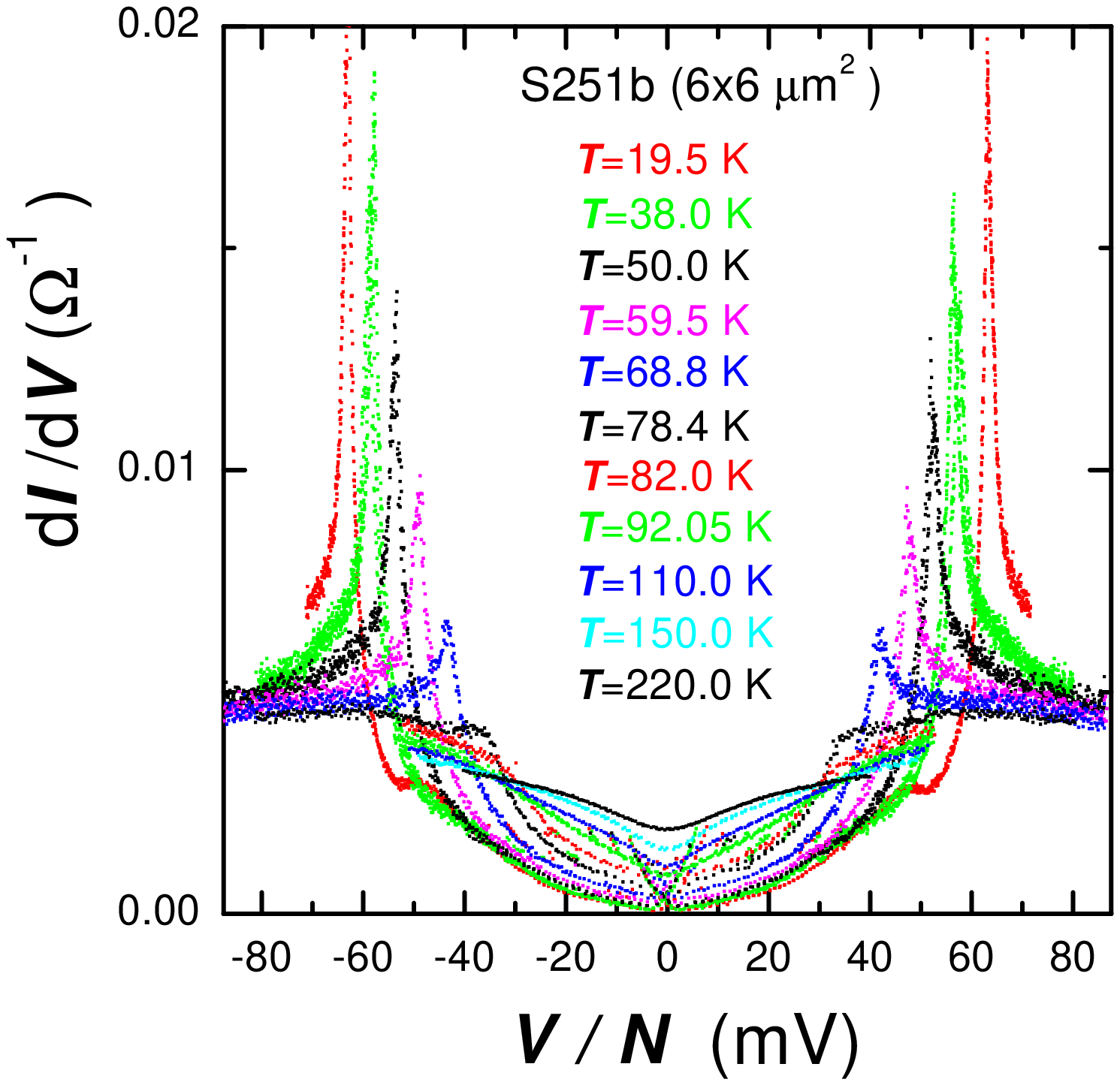,width=0.9\columnwidth}
\caption{Conductance at different temperatures for $6 \times 6$
$\mu$m$^2$ mesa.} \label{fig2}
\end{figure}

There is a crucial difference between our, ''intrinsic'',
and the surface tunneling experiments\cite{Renner,Wilde}, which
allows us to distinguish the superconducting gap from the pseudo
gap. The difference is in existence of multiple quasiparticle
branches in IVC's, see Fig.1. Therefore, we have an additional
quantity, the spacing between the quasiparticle branches, $\Delta
V$, which can be used for estimation of $\Delta $. Even though the
peak in conductance is smeared out
at $T>$80 K, the quasiparticle branches in the IVC's remain well defined up to $%
T_c$. The $\Delta V$ continuously decreases with increasing $T$
and vanishes exactly at $T_c$, as shown in inset a) in Fig.1.

This brings us to conclusion that the superconducting gap does
close at $T_c$, in contrast to the statement of Ref.\cite{Renner}.
On the other hand, the pseudo gap is almost
independent of $T$, in agreement with \cite {Renner,Wilde}. The pseudo
gap can exist well above $T_c$ and, probably, can coexist with the
superconducting gap even at $T<T_c$, see Fig.2. We can not
conclude that the superconducting gap is developed from the pseudo
gap nor that they are competing with each other. From our
experiment, it seems more natural to assume that those two gaps
are independent or only weakly dependent, despite having the same
order of magnitude. In Ref. \cite{Halbritt} possible scenarios of
the pseudo gap were reviewed. One of the possible mechanisms is
due to Coulomb charging effect in IJJ's. Some experimental
evidence for that was obtained in \cite{Latysh}. For the smallest
mesas, we have also seen certain features, such as a complete
suppression of the critical current and an offset voltage in
IVC's, which may be explained in terms of the Coulomb charging
effect. However, further study is necessary before making a
decisive conclusion about the origin of the pseudo-gap.

\end{document}